# Radiofrequency plasma spheroidization of lightweight complex concentrated superalloy $Al_{10}Nb_{20}Ta_{15}Ti_{30}V_5Zr_{20}$

František Lukáč [1], Jakub Klečka [1], Jiří Kozlík [2], Kristína Bartha [2], Jozef Veselý[1], Kateřina Ficková [2], Jakub Minařík[1], Jan Čížek[1], Josef Stráský [2]

[1]Institute of Plasma Physics of the Czech Academy of Sciences, U Slovanky 2525/1a, 182 00, Prague, Czechia

[2]Charles University, Faculty of Mathematics and Physics, Department of Physics of Materials, Ke Karlovu 2026/5, 121 16, Prague, Czechia

email: lukac@ipp.cas.cz

## Abstract

Recent advances in additive manufacturing (AM) have significantly increased the demand for high-quality metal powders with tailored composition. For refractory complex concentrated alloys (RCCA) preparing such powders by conventional metallurgical method remains challenging due to their high melting point and chemical reactivity. In this paper, we present a method to produce spherical RCCA powders (here $Al_{10}Nb_{20}Ta_{15}Ti_{30}V_5Zr_{20}$) via mechanical alloying followed by radio-frequency inductively coupled plasma spheroidization. The resulting powder is suitable for AM processes because of its spherical morphology and a microstructure consisting of major BCC phases with local B2 phase. Complex carbide formation was studied by 3D atom probe tomography.

### Keywords

Refractory complex concentrated alloy, plasma spheroidization, additive manufacturing

## Introduction

The quest for next-generation high-temperature aerospace structural materials is primarily driven by the need to exceed the operational limits of current nickel-based superalloys, which are fundamentally constrained by their melting points and lose structural integrity at temperatures approaching 1200 °C [1,2]. In this context, the development of refractory complex concentrated alloys (RCCAs) represents a paradigm shift in alloy design, utilizing multi-principal elements to achieve exceptional strength and stability at ultra-high temperatures [3–5].

A particularly compelling candidate for these extreme environments is the alloy $Al_{10}Nb_{20}Ta_{16}Ti_{30}V_4Zr_{20}$, originally investigated by Senkov [6] and later studied in the slightly modified $Al_{10}Nb_{20}Ta_{15}Ti_{30}V_5Zr_{20}$ composition [7]. This alloy system was classified as a refractory high-entropy superalloy, characterized by a unique hierarchical nanostructure of coherent ordered B2 and disordered BCC (A2) phases [1,7] This microstructure is a morphological analog to the γ / γ' structure that empowers nickel-based superalloys, yet it is built from refractory elements with significantly higher melting temperatures [2,7].

Furthermore, this alloy addresses the critical "density penalty" associated with early-generation RCCAs. By incorporating lighter elements like aluminum and titanium, the density is reduced to approximately 7.4 g/cm$^3$, roughly 15% lower than typical nickel superalloys, while maintaining superior specific yield strength at elevated temperatures [6,7]. While early iterations suffered from room-temperature brittleness due to a continuous B2 matrix, landmark research into "phase inversion" has demonstrated that controlled thermal treatments can replace the brittle B2 matrix with a ductile BCC matrix, increasing compressive ductility from 4.5% to over 20% [8,9]. Consequently, $Al_{10}Nb_{20}Ta_{15}Ti_{30}V_5Zr_{20}$ served as a key platform for studying strengthening mechanisms, such as grain-boundary pinning by in situ oxides and lattice-distortion effects, which are essential for producing advanced, high-performance structural materials [10,11].

The emergence of complex concentration alloys was also addressed in the field of additive manufacturing (AM) [12]. Here, some of these alloys with dual-phase nanolamellar eutectic microstructure highly outperform 3D-printed traditional alloys in terms of the combination of strength and ductility [13]. However, the industrial scaling of these materials via AM is currently limited by the lack of high-purity, spherical feedstock with a narrow particle size distribution [14,15]. Conventional synthesis routes, such as mechanical alloying (MA), inherently produce irregular, angular, or flaky powder morphologies [16]. These irregular particles have poor flowability and low tap density, leading to stochastic defects and porosity in powder-bed AM processes like selective laser melting (SLM) [17,18].

Radio-frequency inductively coupled plasma (RF-ICP) spheroidization offers the potential to fabricate such high-quality powder materials. Irregular precursor powders are injected axially into the high-temperature plasma jet zone (10,000 K [19,20]), where they undergo near-instantaneous in-flight melting. Once molten, the individual particles adopt a homogeneous chemical composition, and the liquid metal droplets' high surface tension forces them into spherical geometry. Upon exiting the hot zone, particles undergo rapid cooling at rates of approximately $10^6$ K/s, locking the spherical morphology while suppressing chemical segregation [21]. For RCCAs involving heavy elements like tungsten, molybdenum, and tantalum, RF plasma spheroidization is the premier choice because its focused high enthalpy plasma provides the necessary heat input to melt these high-melting-point metals completely [22].

A key motivation for using the RF variant of plasma technology with its electrode-free nature ensures a high-purity, contamination-free environment [23]. This is a major advantage over direct current plasma torches, which often suffer from electrode erosion that can introduce impurities into the alloy system [17]. The RF-ICP also acts as an active refining process where existing surface oxides may be volatilized. Further, introducing hydrogen into the sheath gas acts as a potent reducing agent, significantly reducing oxygen and carbon content in systems like Nb5W2Mo1Zr and WTaMoNbZr, dramatically improving the ductility of the final parts [17,24].

In addition to its advantages over other plasma spray processes, RF-ICP spheroidization is also superior alternative method for synthesizing tailored high-purity refractory powders. For instance, direct methods such as electrode induction melting gas atomization (EIGA), which can be limited in accessibility due to high equipment costs and the energy-intensive nature of creating weight-limited pre-alloyed ingots [22,25], RF spheroidization allows for the continuous processing of accessible irregular precursors into high-quality spherical powders [21]. The plasma rotating electrode process (PREP produces high-sphericity particles with smooth surfaces, but it is constrained by the mechanical limits of electrode rotation speeds and suffers from a coarse particle size distribution and a notably low yield of fine powders required for SLM [17,26]. A key characteristic of plasma spheroidization methods is its ability to preserve the volumetric particle size distribution of the feedstock powder. Furthermore, melt-based atomization methods like ultrasound-atomization use vacuum arc-melt ingots that, in the case of RCCAs, are often susceptible to micro-segregation and solute gradients, leading to inhomogeneous powder batches [27,28].

Successful implementation of RF-ICP spheroidization has been demonstrated on commercial refractory alloys. A prime example is the production of spherical Nb-5W-2Mo-1Zr alloy powder, where process optimization achieved a spheroidization ratio of 98% (i.e. the ratio spheroidized powder particles in output powder) and significantly reduced oxygen and hydrogen interstitials, enabling the fabrication of SLM components with a relative density of 99.81% and an excellent strength-ductility balance [17]. Similarly, Xia et al. successfully synthetized the hydrogenated-dehydrogenated RCCA WTaMoNbZr into highly spherical feedstock with a 95.3% spheroidization ratio, exhibiting a single-phase body-centered cubic (BCC) structure and high nano-hardness, making it ideal for AM applications [24]. Other RCCA, WMoNbTa [29] and VNbMoTaW [30], were produced by RF-ICP spheroidization of spray-granulated powders and consolidation by laser bed powder fusion resulting in good mechanical properties without significant interstitial contamination. This confirms the capability of RF plasma to produce morphologically and chemically uniform powders for safety-critical AM from irregular or contaminated feedstocks after MA [21,23] . Thus, RCCA $Al_{10}Nb_{20}Ta_{15}Ti_{30}V_5Zr_{20}$ prepared by mechanical alloying from elemental powders was spheroidized with RF-ICP for the first time in this work.

The combination of MA and subsequent RF-ICP spheroidization represents a robust pathway for synthesizing RCCA powders that are otherwise difficult to produce via traditional melting routes. By collapsing densification, homogenization, and purification into a single in-flight process, RF-ICP technology enables the fabrication of high-performance alloys that meet the most stringent industrial standards by delivering advanced feedstock with high apparent density, enhanced flowability, and atomic-scale chemical homogeneity.

# Experimental methods

## Material preparation

The input powders were purchased from Thermo Scientific Chemicals (USA) and Camex (Czechia) and their properties are listed in Table 1.

*Table 1 Properties of raw powders. The O and N measurement data were not available for Ta powder*

| | Manufacturer | Purity (%) | Size range (μm) | Oxygen (wt. %) | Nitrogen (wt. %) |
|---|---|---|---|---|---|
| **Al** | Thermo Scientific Chemicals | 99.5 | 7–15 | 0.46±0.01 | 0.00±0.00 |
| **Nb** | Camex | 99.9 | 20–53 | 0.01±0.00 | 0.01±0.00 |
| **Ta** | Thermo Scientific Chemicals | 99.9 | 0–44 | — | — |
| **Ti** | Thermo Scientific Chemicals | 99.5 | 20–53 | 0.39±0.01 | 0.01±0.00 |
| **V** | Camex | 99.9 | 20–53 | 0.19±0.01 | 0.02±0.01 |
| **Zr** | Camex | 96.7* | 20–53 | 0.19±0.01 | 0.02±0.01 |

**2.9% of Hf is a natural impurity in Zr because they are difficult to separate chemically.*

Particle size distribution was measured using a Mastersizer 3000 laser diffraction analyzer (Malvern Panalytical, UK). Wet cell and ultrasound excitation were used to disrupt potential agglomerates.

Mechanical alloying was performed by high-energy ball milling using a planetary mill Pulverisette 5 (Fritsch, Germany). The mixture of powders was inserted into a WC-Co milling jar filled with WC-Co balls with diameter of 10 mm sustaining ball-to-powder ratio of 7:1. Powder was manipulated inside a glovebox with argon atmosphere (GS Glovebox-Systemtechnik, Germany) and pre-milled at 120 rpm for 0.5 h. The mechanical alloying was then performed at 240 rpm for 48 hours. After milling, direct powder collection was not possible, because of coagulation. Instead, we collected it in six subsequent extraction steps; in each, we added a process-controlling agent (0.25 ml ethanol, p.a. purity) and repeated the same milling program for a total of 100 minutes. During each such step, the powder was sieved to 20–50 μm and size fractions out of this range were returned to mill to increase process effectivity.

The MA powder was then spheroidized using an RF-ICP. TekSpray-15 system (TEKNA, Canada) was used at a power of 15 kW and a chamber pressure of 103.4 kPa. The main process gas was argon, with the flows of central gas, sheath gas, carrier gas, and nozzle gas of 10, 35, 8, and 8 slpm, respectively. A small hydrogen flow of 3.8 slpm was added to the sheath gas. The powder was fed into the plasma using a Twin-System feeder (Plasma-Technik AG, Switzerland) at an approximate feed rate of 14.3 g/min (around 860 g/h).

### Experimental characterization

Quantitative phase composition was obtained by X-ray diffraction (XRD) using Bruker D8 Discover (Bruker AXS, Germany) equipped with Cu anode. The data were processed by Rietveld fitting using Topas V5 software. Chemical composition was measured by an XRF analyzer Bruker S2 Puma (Bruker AXS, Germany). The content of oxygen and nitrogen was determined using carrier-gas hot extraction (CGHE) method with a Bruker G8 Galileo.

The powder microstructures as well as free surfaces were observed by scanning electron microscopy (SEM, Apreo 2S, Thermo Fisher Scientific, Czechia, and EVO MA 15 (Carl Zeiss SMT, Germany). The elemental composition was measured by energy-dispersive X-ray spectroscopy (EDS) analyses EDAX Octane Elite EDS detector. Vickers microhardness was measured at 10 gf (HV0.01) and averaged over more than 25 particles for each condition.

Zeiss Auriga Compact Crossbeam $Ga^{+}$ FIB-SEM (Carl Zeiss SMT, Germany) was utilized for transmission electron microscopy (TEM) lamella preparation. The TEM was performed using a JEOL 2200FS microscope (JEOL Ltd., Japan) operated at 200 kV.

For atomic-resolution observations, 3D atom probe tomography (APT) needles were prepared using a ThermoFisher Scios 2 Dualbeam FIB-SEM (Thermo Scientific, Czechia). The analysis was performed at 50 K using CAMECA LEAP 6000 XR (CAMECA, USA) in laser mode, using a pulse energy of 40 pJ at 200 kHz with 0.3 % target detection efficiency. The reconstruction and post-processing were performed in IVAS 6.3 software (CAMECA, USA).

### Theoretical calculations

Calculation of melting and boiling point of the alloy was performed in Thermocalc using TCHEA8 database (Thermocalc, Sweden).

## Results and discussion

### Microstructure, chemical composition and hardness

Figure 1 shows typical $Al_{10}Nb_{20}Ta_{15}Ti_{30}V_5Zr_{20}$ powder particles after 48 hours of the mechanical alloying process. The MA particles are irregularly shaped, but not heavily fragmented, a morphology typical of the MA process before complete depletion of plastic deformation [31] . The cross-section shows a relatively even composition, homogeneous at grain level. The infrequent bright inclusions are WC-Co fragments that spalled from the milling jar and balls due to process intensity. Their overall content is responsible for 1.81 wt. % (0.86 at. %) of tungsten in the powder as shown in Table 2. In addition, the MA process caused only insignificant changes in the relative ratios of the six elements. The oxygen and nitrogen content in the MA powder matched the values in the original feedstock powders and remained within acceptable limits, see Table 3. The gas impurities should be avoided since they are responsible for significant solid solution strengthening and formation of hard precipitates greatly influencing mechanical properties of similar alloys (nitrides TiN [32], oxides Mn-Si-O [33], oxy-nitride complexes TiHfZr-ON [34]). Measured particle size distribution (PSD) was adjusted to be ideal for additive manufacturing methods by sieving during extraction process after milling, with $D_{50}$ value of 37.3 μm. This approach provides narrow PSD for the RF-ICP process parameters optimization and avoids material loss during final sieving for AM methods, which is inevitable in pulverization by hydrogenation-dehydrogenation method [35]. The mechanical alloying yielded 40.14 g of powder with desired particle size distribution from 50 g of feedstock, resulting in 80 wt.% yield.

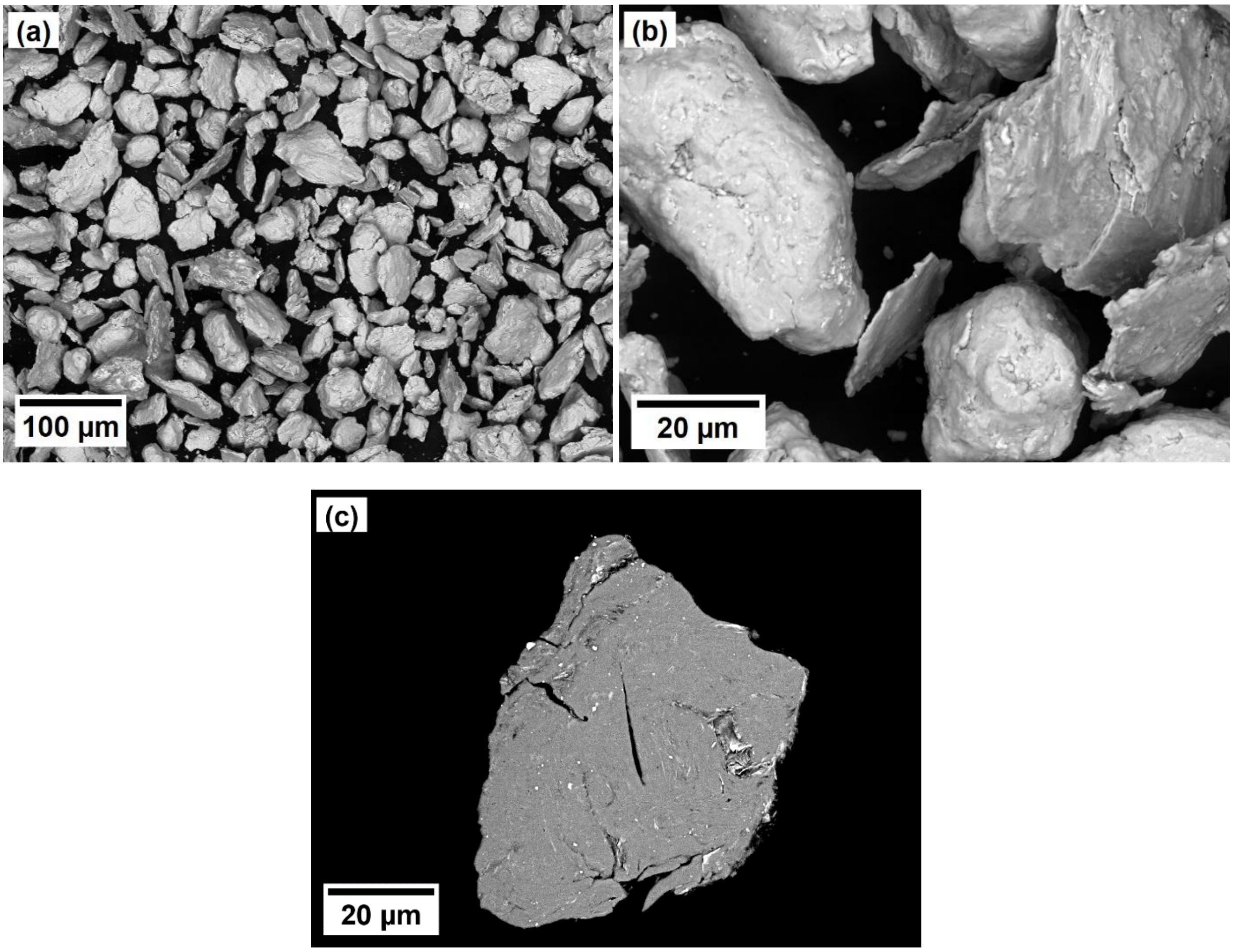

*Figure 1: Morphology and microstructure of the MA $Al_{10}Nb_{20}Ta_{15}Ti_{30}V_5Zr_{20}$ powder by SEM a), b) overall appearance (SE signal), c) cross section (BSE signal),*

*Table 2: Chemical composition of the MA and SP powders (in wt.%) determined by XRF (metals), CGHE (oxygen and nitrogen). Powders acquired from the container (SP-C) and wall (SP-W) of the atomization chamber were separated for better comparison.*

| | Nominal | MA | SP-C | SP-W |
|---|---|---|---|---|
| **Al** | 3.22 | 2.49 | 1.77 | 6.59 |
| **Nb** | 22.24 | 21.87 | 20.70 | 17.31 |
| **Ta** | 32.48 | 34.51 | 38.31 | 36.03 |
| **Ti** | 17.18 | 15.12 | 13.29 | 16.78 |
| **V** | 3.04 | 2.91 | 2.68 | 2.87 |
| **Zr** | 21.84 | 20.69 | 20.56 | 17.92 |
| **Hf** | | 0.53 | 0.54 | 0.50 |
| **W** | | 1.81 | 2.11 | 1.95 |
| **O** | | 0.40±0.00 | 0.36±0.01 | 0.66±0.14 |
| **N** | | 0.02±0.00 | 0.10±0.01 | 0.22±0.06 |

*Table 3: Parameters of the particle size distribution and efficiency of the processing steps for MA powder and powders extracted from spheroidization chamber bottom container (SP-C) and walls (SP-W).*

| powder | $D_{10}$ (μm) | $D_{50}$ (μm) | $D_{90}$ (μm) | Yield (wt. %) |
|---|---|---|---|---|
| **MA** | 20.8 | 37.3 | 64.2 | 80 |
| **SP-C** | 26.1 | 39.3 | 58.2 | 15 |
| **SP-W** | 4.0 | 26.8 | 46.2 | 72 |

After spheroidization, the spheroidized powder (SP) collected from the container of the spheroidization chamber (SP-C – 15 wt.% of feedstock) has particle size distribution of $D_{50}$ = 39.3 μm and the powder collected from the walls of the spheroidization chamber (SP-W – 72 wt.% of feedstock) as particle size distribution of $D_{50}$ = 26.8 μm, as presented in Table 3. This gives overall yield of the process (milling + spheroidization) as 70 wt.%.

In our spheroidization process, a maximum TekSpray-15 system power of 15 kW was used. This is lower than other reports on spheroidized refractory alloys [19,36]. Despite that, almost all particles were spheroidized, as illustrated in Figure 2a. Table 3 shows that the particles attached to the wall are smaller in size than particles which fall to the bottom of the container. Figure 2b shows the SP-W powder, which exhibits low number of satellites typical for spheroidized powder; these are submicron particles embedded in the surface of larger particles resulting from in-flight collisions [37]. During the spheroidization process, some material evaporated and was condensed at the surface of the particles already attached to the wall of the spheroidization chamber, appearing as a darker fluffy layer, visible in Figure 2b and c. This is a result of evaporation of elements which exhibit higher vapor pressure at the elevated temperatures, see Table 4 – the predicted alloy vapor contains roughly 91 at.% of Al and 9 at.% of Ti. During the flight through the hot zone of plasma, smaller particles can be overheated and aluminum evaporation occurs. This is documented by higher-than-nominal aluminum content in the powder collected from the wall of the spheroidization chamber measured by XRF and presented in Table 1. The XRF is mainly detecting excess aluminum at the surface of measured powder particles and thus overestimating the actual Al content in the whole powder volume.

A particularly high nitrogen concentration was detected after RF-ICP treatment in particles adhering to the chamber wall (Table 2). This may be attributed to the high-temperature processing of powders rich in Ti and Zr, which are highly effective gas getters and have a strong affinity for nitrogen. Nitrogen source likely originated from residual air within the powder feeder, despite extensive argon gas purging, high flow rates, and positive system pressure. This potential source of contamination should be eliminated by using closed feeder filled in glove box with protective atmosphere rather than only purging the powder feeder with argon, particularly when processing RCCAs containing elements prone to nitridation.

*Table 4: Melting points, boiling points, and vapor pressure [38] of the individual elements, and alloy estimates by rule of mixture and by ThermoCalc (TCHEA8 database) for the nominal alloy composition. According to CALPHAD, the vapors of the alloy consist of approximately 91 at.% of Al and 9 at.% of Ti.*

| Composition | Melting point (°C) | Boiling point (°C) | Vapor pressure at 2000 °C (Pa) |
|---|---|---|---|
| Al | 660 | 2470 | $7.7 \times 10^{3}$ |
| Ti | 1668 | 3287 | $3.2 \times 10^{1}$ |
| V | 1910 | 3407 | $1.3 \times 10^{1}$ |
| Nb | 2477 | 4744 | $1.9 \times 10^{-4}$ |
| Zr | 1852 | 4377 | $1.2 \times 10^{-2}$ |
| Ta | 3017 | 5458 | $4.1 \times 10^{-6}$ |
| Rule of mixture | *1980* | *4046* | — |
| CALPHAD | *1468–1953* | *3203–4972* | $8.6 \times 10^{1}$ |

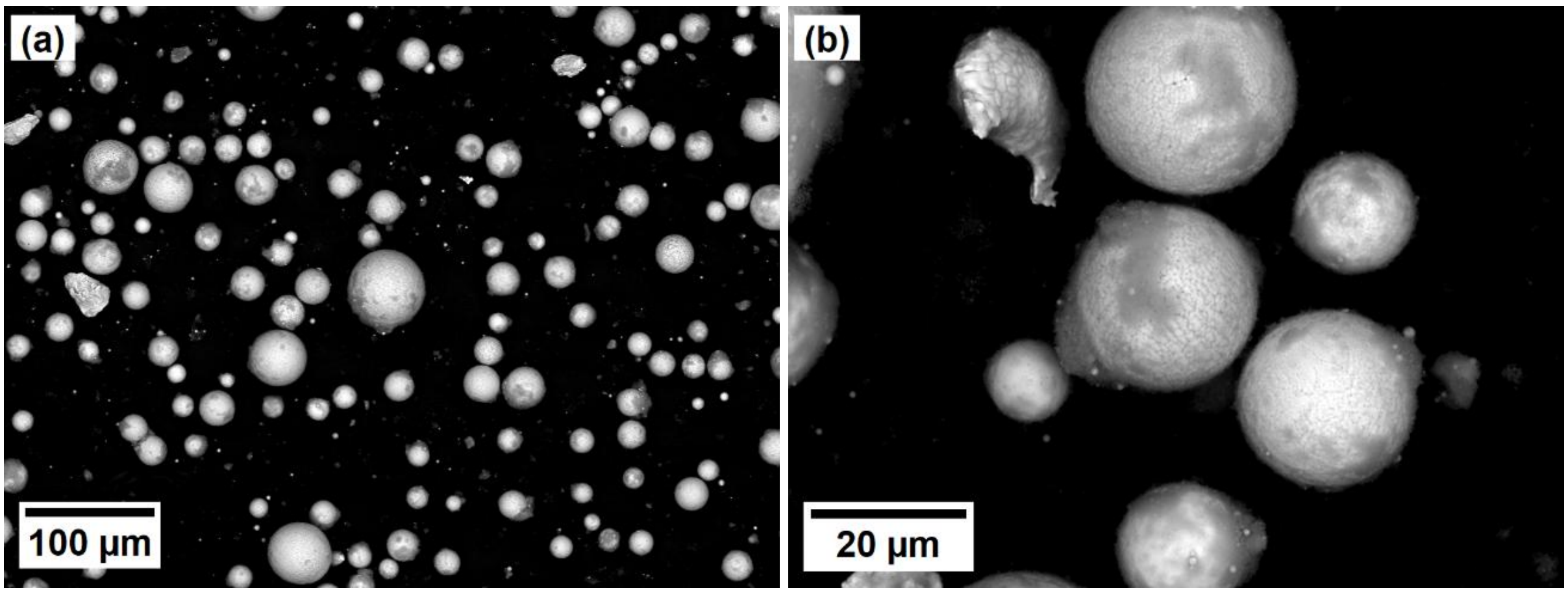

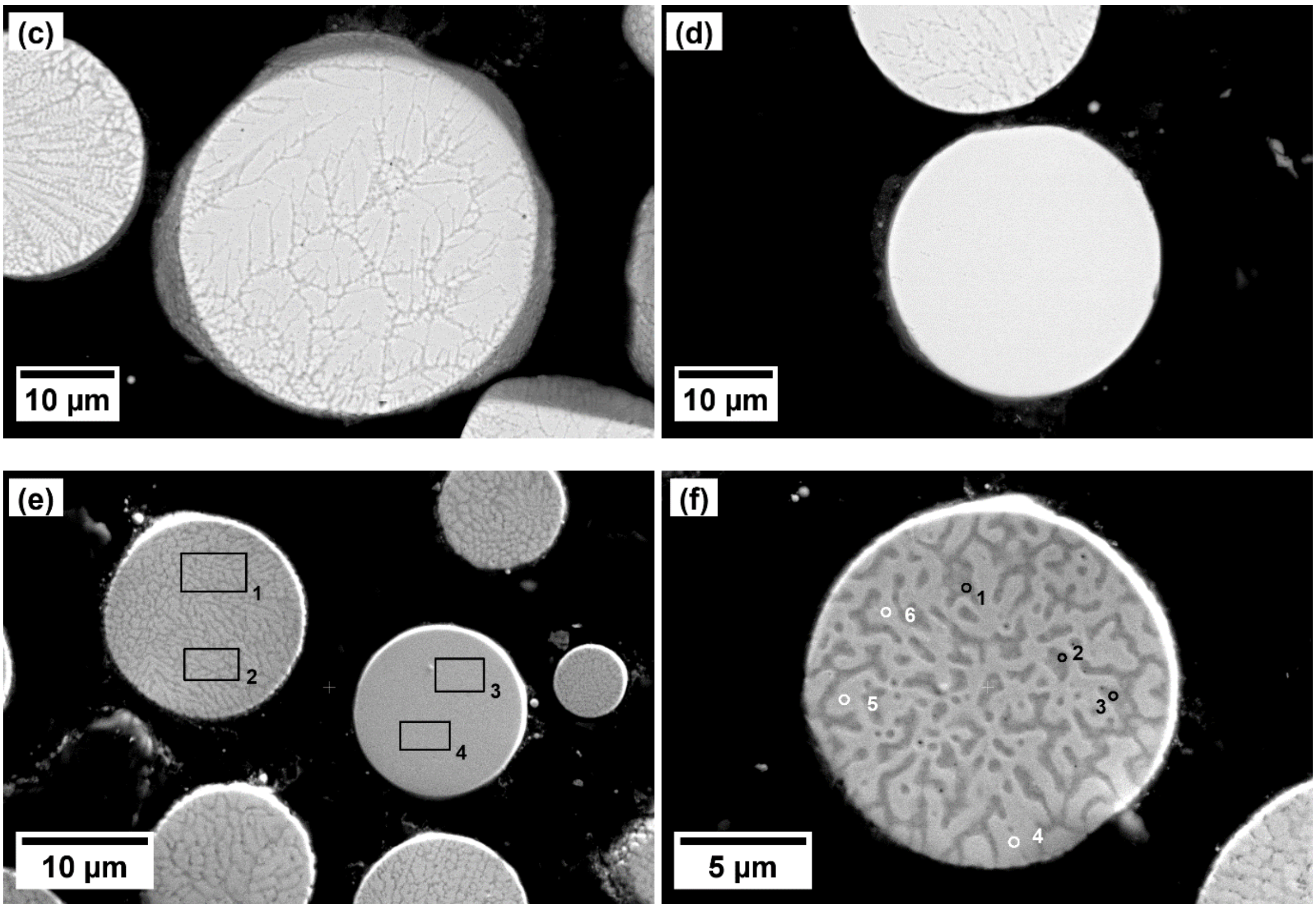

*Figure 2: Morphology and microstructure of the SP $Al_{10}Nb_{20}Ta_{15}Ti_{30}V_5Zr_{20}$ powder; a), b) free powder, c) cross-section of the dual-phase particle, d) cross-section of the single-phase particle, e) and f) cross sections with areas of the EDS analysis marked, results can be found in Table 5.*

As shown in Figure 2c-f, two distinct microstructures were obtained during spheroidization: dendritic, dual-phase particles and much less frequent single-phase particles. The averaged compositions obtained by EDS provided in Table 5 suggests that two phases in the dual-phase material differ significantly: while the inter-dendritic regions are enriched in Zr and Al, the dendritic grains were enriched in Nb and Ta. Senkov et al. [7] investigated two constituent phases formed during casting of the $Al_{10}Nb_{20}Ta_{15}Ti_{30}V_5Zr_{20}$ alloy and reported similar chemical compositions after homogenization annealing; BCC $Al_{4.4}Nb_{21}Ta_{22.5}Ti_{33.0}V_7Zr_{12.1}$ (Ta enriched) and B2 $Al_{16.2}Nb_{10.4}Ta_{5.5}Ti_{32.9}V_{4.5}Zr_{30.6}$ (Al, Zr enriched).

The chemical composition of the dual-phase and single-phase particles is shown in Table 5. Dual-phase particles have a chemical composition which is very close to the nominal composition of the alloy, while the single-phase particles are depleted of Al and Ti. These are the elements with the lowest boiling points, suggesting such particles overheated during RF plasma spheroidization. They are less frequent in the collected spheroidized powder, and their size is comparable to dual-phase particles, thus no correlation of these overheated particles with their size is found.

*Table 5 The chemical composition of dual-phase and single-phase particles in Fig 2e, the chemical composition of both phases in Fig 2f, where spots 1-3 are averaged as dark phase, spots 4-6 are averaged as light phase, in at. %.*

| | Al | Ti | V | Zr | Nb | Ta |
|---|---|---|---|---|---|---|
| nominal | 10.0 | 30.0 | 5.0 | 20.0 | 20.0 | 15.0 |
| dual-phase particle | 10.3 | 33.4 | 5.9 | 19.1 | 18.6 | 12.7 |
| single-phase particle | 3.7 | 30.7 | 6.4 | 22.2 | 22.3 | 14.6 |
| dark phase | 9.3 | 34.4 | 7.3 | 28.8 | 13.6 | 6.6 |
| light phase | 4.6 | 32.3 | 5.7 | 14.3 | 25.3 | 17.9 |

Microhardness of powder particles was measured selectively for single-phase particle as 715 ± 119 HV0.01 and dual-phase particle as 710 ± 87 HV0.01. In other words, despite the different microstructure and chemical composition, the hardness of the two particle types remained almost identical. At the same time, these values are significantly higher than the hardness of (530 HV [6] of a corresponding ingot which exhibited very fine dual-phase microstructure analogous to our dual-phase particles after its long-term annealing at 1200 °C and slow cooling. The high microhardness may be attributed to high nitrogen level in the spheroidized powders, causing strong interstitial strengthening, which overcomes the effects of phase composition.

## Phase composition

Figure 3 shows XRD patterns of the MA and SP powders with the BCC structure as the major phase. The lattice parameter of phase in MA powder was 0.3327 nm. The small content of the WC impurities observed in the MA powder by XRF was confirmed by the XRD and indicated as 2.9 wt. %.

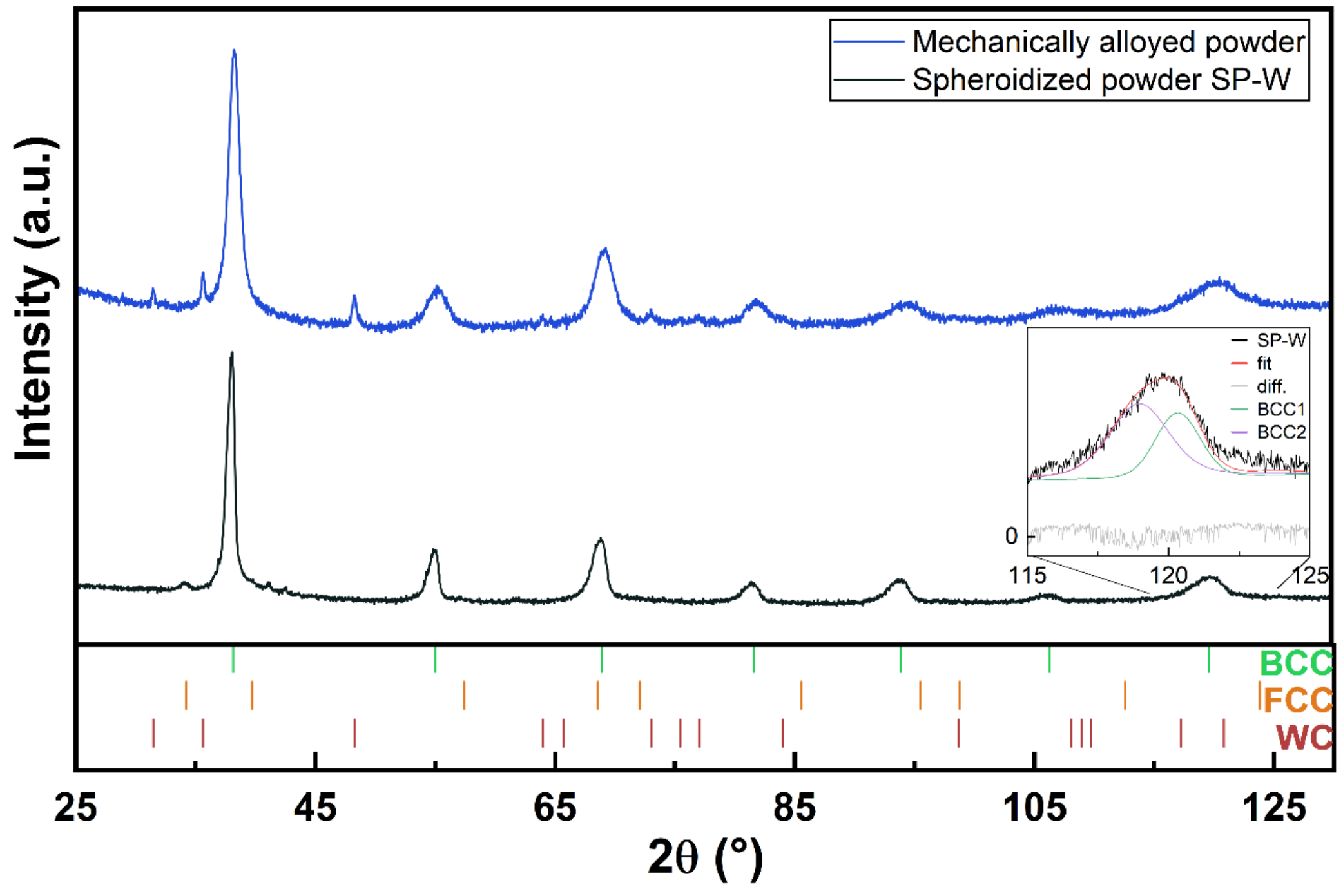

*Figure 3: XRD patterns of the mechanically alloyed powder and RF plasma spheroidized powder collected from the wall of spheroidizing chamber.*

The SP-W powder exhibited two BCC phases. The first, identical with that observed in the MA powder, with the lattice parameter of $a_1$ = 0.3327 nm (63 wt.%), and a second BCC phase with a slightly different lattice parameter of $a_2$ = 0.3357 nm (35 wt.%), see inset of Figure 3. These values are slightly higher than the values of $a_1$ = 0.3301 nm and $a_2$ = 0.3320 nm reported by Senkov et al. [6] after casting and 1200 °C / 24 h annealing. The lattice parameter of the secondary BCC phase is close to the value of 0.3360 nm calculated by Brodie et al. for the B2 phase observed in this alloy [39]. No additional peaks of an ordered B2 phase were observed in XRD due to the fact that these low intensity peaks are under the limit of detection when B2 phase content is low. The WC phase was not detected after the RF spheroidization process; the high-temperature thermal treatment most likely led to the dissolution of the WC impurities. On the other hand, a new FCC phase with a lattice parameter of 0.455 nm emerged after spheroidization, corresponding to zirconium-rich carbide particles of ZrC structure. This compound is likely formed due to the high affinity of Zr and C; here, the carbon originated from the ethanol added as a process control agent, and from the dissolved WC particles.

## TEM

Following the phase composition results, TEM analysis of the double-phase SP-C particle was performed to distinguish between the possible BCC and B2 phases and further analyze the ZrC-based phase. The distribution of ordered and disordered phase is shown in Figure 4b. As seen in Figure 4a, b and e, the B2 phase is present in the inter-dendritic Zr-rich regions (marked with red circles), while the Ta- and Nb-rich regions are disordered. Ti and V seem to be equally distributed between the two regions. The B2-exclusive peaks are significantly weaker than the BCC peaks, explaining absence of this phase in the XRD measurement above. A possible reason for this may be high cooling rates during RF-ICP spheroidization, allowing only weak ordering of elements between the two B2 sublattices, since well-developed B2 phase was observed after long-term annealing at 1200 °C [40]. However, the low B2 peak intensity may also be a result of a particular combination of elements at the two sublattices, yielding similar average atomic scattering factors for both sites.

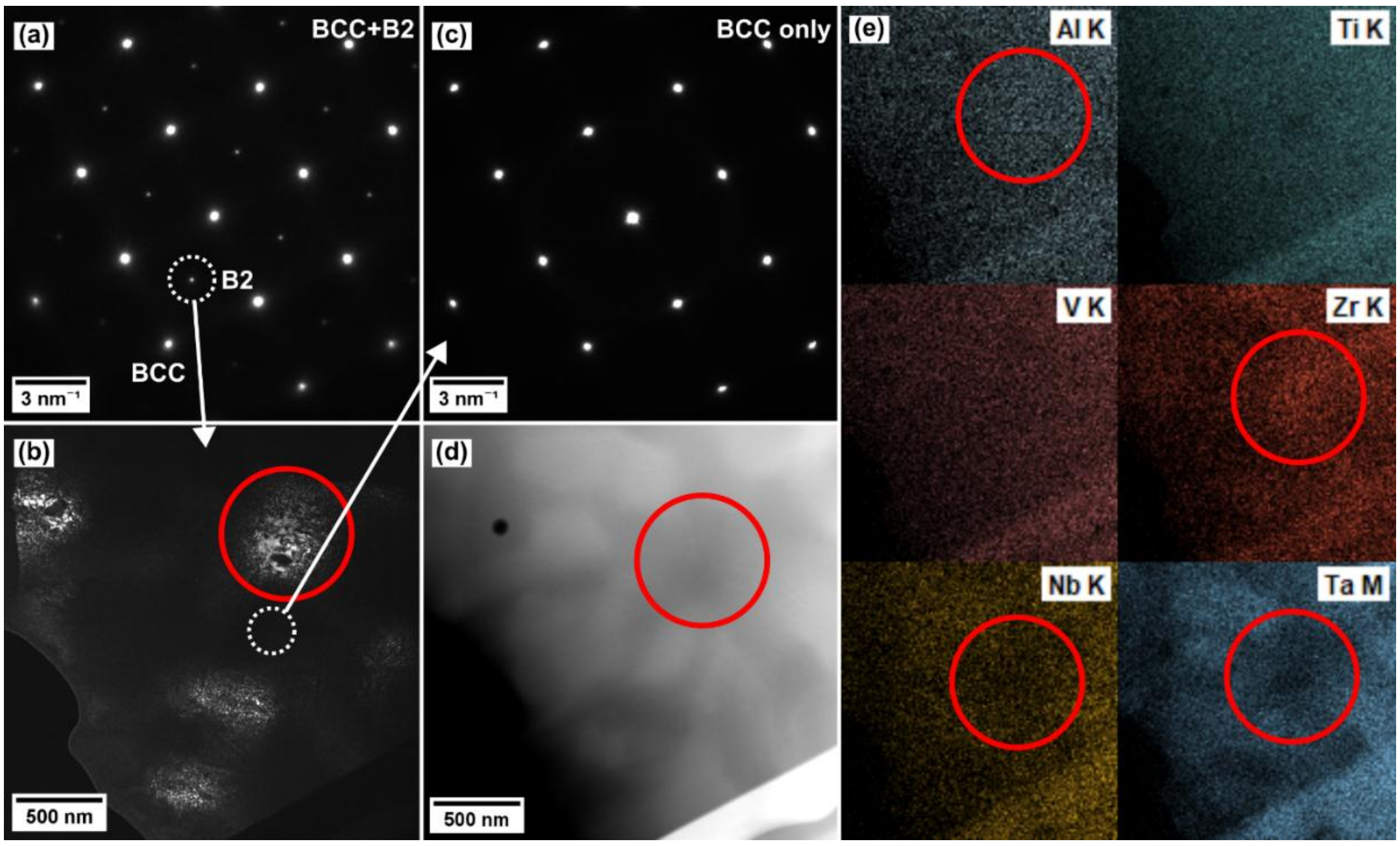

*Figure 4: TEM analysis of the SP powder particle. (a) Electron diffraction pattern showing both BCC and B2 diffraction spots. (b) Dark-field micrograph using the B2 diffraction spot marked in (a), showing the spatial inhomogeneity of the B2 phase distribution. The absence of B2 diffraction between these regions is confirmed by the SAED shown in (c). Finally, STEM micrograph (d) and EDS elemental maps (e) are presented for the same area. The B2 regions correspond to Zr- and Al-rich regions (marked with red circles), while the BCC-only regions exhibit higher concentration of Nb and Ta.*

No other intermetallic precipitates were observed, but the presence of ZrC was confirmed by TEM, as presented in Figure 5. From the detail shown in Figure 5c, the precipitates are rather small, typically hundreds of nm in length and tens of nm in width. The $(111)_{ZrC}$ plane is aligned with the $(011)_{BCC/B2}$ planes. Such orientation relationship is in line with the predicted relationship for W(bcc)-ZrC interfaces in [41].

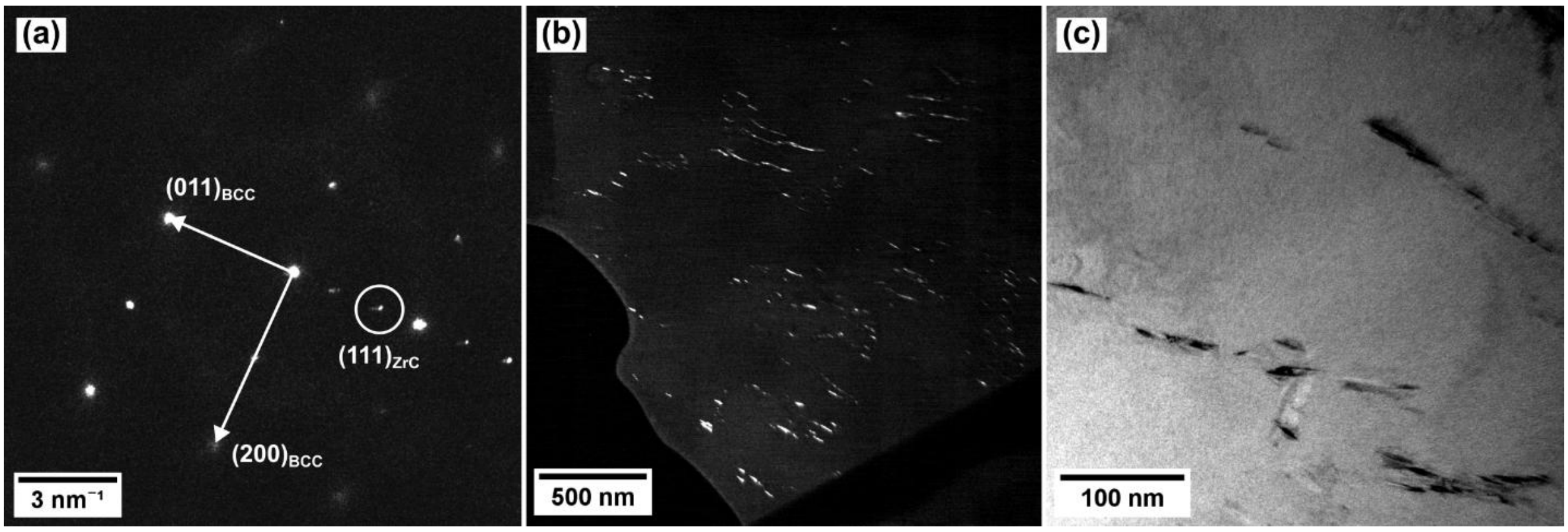

*Figure 5: TEM observations of the ZrC particles. (a) Electron diffraction pattern showing the ZrC diffraction spots. (b) Dark-field micrograph showing the ZrC distribution within the region observed in Figure 4. (c) Detailed view of the ZrC precipitation.*

## 3D APT

Finally, 3D APT was performed to confirm the presence of complex Zr-rich carbides also chemically. The analysis results are presented in Figure 6. Approximately 4 nm thick carbon-rich plates were detected in the sample, as shown in Figure 6a. The concentration profile taken across the plate (marked with cyan cylinder) shows that the plate is rich in both carbon and zirconium, while still containing traces of other metallic elements, finally confirming the nature of the precipitates. The bulk composition analysis far from the ZrC precipitate yields 29.1 at% of Ti, 22.8 at.% of Nb, 17.8 at.% of Ta, 14.0 at.% of Zr, 8.6 at.% of Al and 5.3 at.% of V (the rest belonging to impurities such as O, C etc.). While the analysis of light element impurities in APT should be taken cautiously, the presence of contamination by process control agents is very common during MA [33]. Similar phases which influence mechanical properties and phase stability have already been identified as carbides by methods like APT before [42–45].

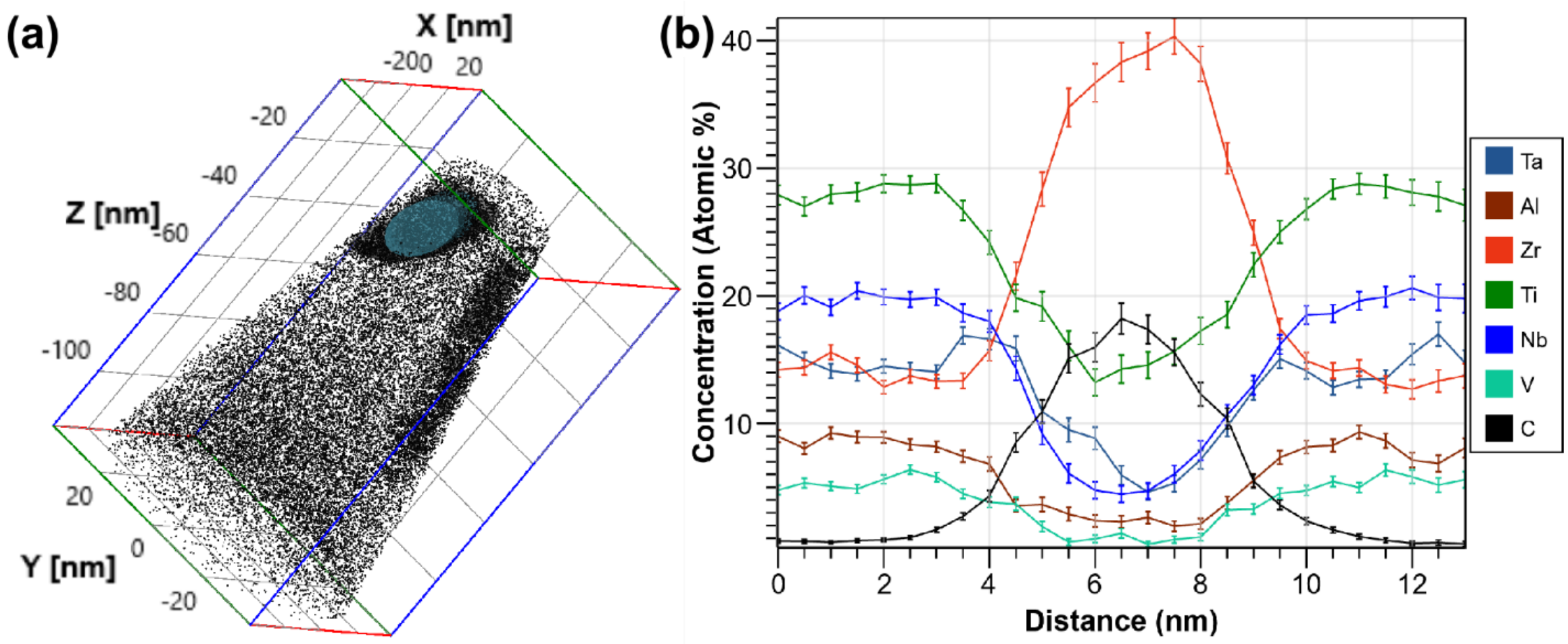

*Figure 6: 3D APT analysis confirming the presence of ZrC precipitates within the matrix. (a) Distribution of C ions within the tip, the 5 at. % isosurface is overlaid. Two precipitates are visible near the tip edges. The elliptical cylinder marks the region, which was used for calculation of the 1D compositional profile. (b) The concentration line profile calculated across the C-rich precipitate along the axis of the cyan cylinder shown in (a). Apart from carbon, Zr concentration is also increased within the precipitate, compared to the matrix.*

## Discussion

Radio-frequency plasma spheroidization of the $Al_{10}Nb_{20}Ta_{15}Ti_{30}V_5Zr_{20}$ alloy provided powder with high degree of sphericity and particle size of approximately $D_{50}$ 33 μm (average of SP-C and SP-W particles). Such powder may be suitable for deposition using additive manufacturing methods. For most common method of selective laser melting (SLM), the powder needs to be spread evenly to achieve even thickness of each layer. Therefore, uniform particle size, packing density and flowability are the key factors for successful builds. Indeed, the prepared powder has high sphericity and appropriate particle size distribution in the range of 10–53 μm typically reported for SLM [45]. As the particles are completely melted during the SLM process, no additional requirements are imposed on the microstructure or phase composition of the powder. The effect of the morphology of the powder on AM process effectiveness should be studied further.

The satellites formation documented in Figure 2b is caused by evaporation of smallest particles of powder and by evaporation of low-melting-point elements of the alloy and their subsequent condensation. This can be suppressed by altering parameters of the process to decrease the exposure of the particle to the plasma by increasing their speed or decreasing the plasma enthalpy. However, as a result, the ratio of spheroidized particles may be sacrificed due to the tradeoff between spheroidization ratio and evaporation, well described in [21].

Despite the successful spheroidization, the processed powder contained significant amount of oxygen and, namely, nitrogen. This observation highlights a critical aspect of RF plasma processing of reactive refractory alloys: morphological improvement does not necessarily coincide with improved chemical purity. Oxygen and nitrogen are interstitial elements in several constituent metals, particularly Ti and Zr, and their incorporation can modify lattice parameters, local chemical environments, phase stability, and mechanical behavior. Nitrogen is well established as a highly effective solid-solution-strengthening element in titanium [46] and it is a possible cause of measured elevated microhardness of spheroidized powder particles. Despite slight overpressure in the system, potential nitrogen sources include residual gas impurities, leakage at seals or connections, moisture-related contamination, gas-line contamination, and powder feeder sealing. For this alloy, where Ti and Zr can act as effective getters for reactive gaseous species, even a low-level impurity source may become relevant under high-temperature processing conditions. This makes the purity of the plasma-forming and carrier gases, as well as the integrity of the chamber and feed system, particularly important. These findings underline a broader methodological point. Oxygen and nitrogen should be measured alongside particle morphology and phase composition when developing RF-PS routes for reactive refractory alloys. When interstitial concentrations are not reported, apparent differences in hardness between studies may be mistakenly attributed entirely to alloy composition, grain size, or phase constitution. This limits the value of inter-study comparisons and may obscure this, potentially decisive, source of variability in mechanical properties.

The presence of varying microstructure of SP particles in Figure 2 showing dual-phase and single-phase is particularly interesting. The reason behind is that during solidification, the tendency to form dual-phase dendritic structures is governed by the alloy freezing range. When Al evaporates extensively, the freezing range gets narrower. According to ThermoCalc calculation, the freezing range is 1908–1461 °C for dual-phase and 1933–1642 °C for single-phase composition (cf. Table 5). The narrower freezing range is also accompanied by smaller difference between solid and liquid phase; it significantly reduces the solidification duration and reduces the space for undercooling. Thanks to that, it may be possible to bypass the dendritic solidification completely.

The consequences of rapid cooling rates on liquid particle solidification were studied in detail. Contrary to other RCCAs containing Al, the stable Zr-Al intermetallic precipitates were not formed thanks to fast thermal treatment during spheroidization [47–49]. Detailed microstructure was studied by XRD of spheroidized powder. However, decisive B2 ordering may be difficult to establish from laboratory XRD when the ordered domains are very small, their volume fraction is limited, or the superlattice reflections are weak relative to the fundamental BCC reflections due to complex element mixing. The absence of clearly resolved B2 reflections in the laboratory XRD pattern should not be interpreted as proof that B2 ordering is absent. However, the cuboidal microstructure of interwoven B2/BCC which is evolving after specific long-term annealing [6,8,9,11] was expectedly not achieved and the alloy powder can be further tailored by suitable thermal treatment of final component. In this work, the B2 phase nuclei were found in the locations enriched in Zr and Al by dendritic segregation and further studies of their evolution during material processing should follow.

Detailed study of real microstructure by TEM and APT shows important findings about the occurrence of carbide impurities. While TEM enables us to see the crystal structure of the nano-sized particle, the APT provided their local chemical composition including the approximate metal-to-carbon ration, which is very demanding on this scale in bulk materials. Possible sources of carbon in our sample come from MA using the process control agent for extraction of cold-welded material. Our trial experiments confirm that the milling of pure Zr powder produced the FCC phase similar to that achieved even after spheroidization, suggesting these ZrC rich carbides are very stable. Similar conclusions were drawn in [33] for transition metal complex concentrated alloys prepared by MA. These findings are nowadays possible to confirm by 3D APT with atomic resolution and emphasize the detailed investigation of basic principles used for decades by modern high-definition methods.

Overall, the results demonstrate that RF plasma spheroidization represents a promising route for the preparation of high-quality RCCA powders, powder reconditioning and resource-efficient alloy processing, while highlighting the need for process optimization and careful control of phase formation and impurity levels to achieve feedstock properties tailored to the targeted additive manufacturing technology. Achieving this potential will require chemical purity and microstructural evolution to be treated as integral parts of process optimization rather than secondary considerations.

## Conclusions

The present study establishes the feasibility of producing predominantly spherical $Al_{10}Nb_{20}Ta_{15}Ti_{30}V_5Zr_{20}$ RCCA powder using RF plasma spheroidization. At the same time, the results identify chemical purity and microstructural characterization as important challenges that must be addressed before the material can be considered fully optimized for AM.

RF-ICP spheroidization successfully transformed the irregular precursor into spherical particles, demonstrating its applicability to this refractory multicomponent system. The detected oxygen and nitrogen concentrations raise concerns regarding contamination during processing and the possible contribution of interstitial strengthening to the measured microhardness. The TEM observations support the coexistence of BCC and B2 regions despite the lack of clearly resolved B2 superlattice peaks in laboratory XRD, illustrating the limitations of relying on a single characterization technique. The observed carbides warrant further analysis to determine their composition, origin and possible influence on the properties of the powder.

Important recommendations arise from this work about facts often neglected in available literature: quantitative control of oxygen and nitrogen through improved gas purity and chamber integrity; detailed characterization of the BCC/B2 phase distribution using electron diffraction and complementary nanoscale chemical analysis; and correlative TEM/APT characterization of impurities originated from MA. Controlled heat treatments could then establish whether the phase morphology can be tailored toward a more clearly defined two-phase architecture, provided that such an architecture is required for the intended application. This study demonstrates process efficiency and highlights its demanding nature by carefully monitoring potential impurities.

## Data availability

The raw data that support the findings of this study are openly available through Zenodo repository at https://dx.doi.org/10.5281/zenodo.23391228 under the CC-BY 4.0 license.

## Declaration of generative AI and AI-assisted technologies in the manuscript preparation process.

Statement: During the preparation of this work, the authors used Gemini 3.6 in order to improve language and structure of the paragraph for readability to native speakers. After using this tool, the authors reviewed and edited the content as needed and take full responsibility for the content of the published article.

## Acknowledgements

This work was financially supported by Czech Science Foundation under Project No. 25-16631S. Financial support by the Operational Programme Johannes Amos Comenius of the MEYS of the Czech Republic, within the frame of the project Ferroic Multifunctionalities (FerrMion) [project No. CZ.02.01.01/00/22_008/0004591], co-funded by the European Union is also gratefully acknowledged.

## CRediT statement

**F.L.**: Conceptualization, Data Curation, Investigation, Methodology, Visualization, Writing – original draft, Project administration, Funding acquisition, Resources; **J.Kl.**: Conceptualization, Investigation, Methodology, Resources; **J.Ko.**: Conceptualization, Data Curation, Investigation, Methodology, Visualization, Funding acquisition, Writing – original draft; **K.B.:** Investigation, Visualisation, Resources; **K.F.**: Investigation; **J.M.**: Investigation; **J.Č.**: Supervision, Conceptualization; **J.S.**: Supervision, Project administration, Funding acquisition; **All:** Writing – review and editing

## References

[1] D.B. Miracle, M.H. Tsai, O.N. Senkov, V. Soni, R. Banerjee, Refractory high entropy superalloys (RSAs), Scr. Mater. 187 (2020) 445–452. https://doi.org/10.1016/j.scriptamat.2020.06.048.

[2] P. Kumar, S.J. Kim, Q. Yu, J. Ell, M. Zhang, Y. Yang, J.Y. Kim, H.K. Park, A.M. Minor, E.S. Park, R.O. Ritchie, Compressive vs. tensile yield and fracture toughness behavior of a body-centered cubic refractory high-entropy superalloy Al0.5Nb1.25Ta1.25TiZr at temperatures from ambient to 1200°C, Acta Mater. 245 (2023) 118620. https://doi.org/10.1016/j.actamat.2022.118620.

[3] D.B. Miracle, O.N. Senkov, A critical review of high entropy alloys and related concepts, Acta Mater. 122 (2017) 448–511. https://doi.org/10.1016/J.ACTAMAT.2016.08.081.

[4] O.N. Senkov, G.B. Wilks, D.B. Miracle, C.P. Chuang, P.K. Liaw, Refractory high-entropy alloys, Intermetallics (Barking). 18 (2010) 1758–1765. https://doi.org/10.1016/J.INTERMET.2010.05.014.

[5] N.R. Philips, M. Carl, N.J. Cunningham, New Opportunities in Refractory Alloys, Metallurgical and Materials Transactions A 2020 51:7 51 (2020) 3299–3310. https://doi.org/10.1007/S11661-020-05803-3.

[6] O.N. Senkov, C. Woodward, D.B. Miracle, Microstructure and Properties of Aluminum-Containing Refractory High-Entropy Alloys, JOM 2014 66:10 66 (2014) 2030–2042. https://doi.org/10.1007/S11837-014-1066-0.

[7] O.N. Senkov, B. Crossman, S.I. Rao, J.P. Couzinie, D.B. Miracle, T.M. Butler, R. Banerjee, M. Mills, Mechanical properties of an Al10Nb20Ta15Ti30V5Zr20 A2/B2 refractory superalloy and its constituent phases, Acta Mater. 254 (2023) 119017. https://doi.org/10.1016/J.ACTAMAT.2023.119017.

[8] V. Soni, B. Gwalani, T. Alam, S. Dasari, Y. Zheng, O.N. Senkov, D. Miracle, R. Banerjee, Phase inversion in a two-phase, BCC+B2, refractory high entropy alloy, Acta Mater. 185 (2020) 89–97. https://doi.org/10.1016/J.ACTAMAT.2019.12.004.

[9] V. Soni, O.N. Senkov, B. Gwalani, D.B. Miracle, R. Banerjee, Microstructural Design for Improving Ductility of An Initially Brittle Refractory High Entropy Alloy, Scientific Reports 2018 8:1 8 (2018) 8816-. https://doi.org/10.1038/s41598-018-27144-3.

[10] L. Moravcikova-Gouvea, I. Moravcik, V. Pouchly, Z. Kovacova, M. Kitzmantel, E. Neubauer, I. Dlouhy, Tailoring a Refractory High Entropy Alloy by Powder Metallurgy Process Optimization, Materials 2021, Vol. 14, Page 5796 14 (2021) 5796. https://doi.org/10.3390/MA14195796.

[11] J.P. Couzinié, M. Heczko, V. Mazánová, O.N. Senkov, M. Ghazisaeidi, R. Banerjee, M.J. Mills, High-temperature deformation mechanisms in a BCC+B2 refractory complex concentrated alloy, Acta Mater. 233 (2022) 117995. https://doi.org/10.1016/J.ACTAMAT.2022.117995.

[12] T. Ron, A. Shirizly, E. Aghion, Additive Manufacturing Technologies of High Entropy Alloys (HEA): Review and Prospects, Materials 2023, Vol. 16, Page 2454 16 (2023) 2454. https://doi.org/10.3390/MA16062454.

[13] J. Ren, Y. Zhang, D. Zhao, Y. Chen, S. Guan, Y. Liu, L. Liu, S. Peng, F. Kong, J.D. Poplawsky, G. Gao, T. Voisin, K. An, Y.M. Wang, K.Y. Xie, T. Zhu, W. Chen, Strong yet ductile nanolamellar high-entropy alloys by additive manufacturing, Nature 2022 608:7921 608 (2022) 62–68. https://doi.org/10.1038/s41586-022-04914-8.

[14] H. Dobbelstein, M. Thiele, E.L. Gurevich, E.P. George, A. Ostendorf, Direct Metal Deposition of Refractory High Entropy Alloy MoNbTaW, Phys. Procedia 83 (2016) 624–633. https://doi.org/10.1016/J.PHPRO.2016.08.065.

[15] A. Mohammadnejad, M. Sun, N.A. Yarlapati, M. Hasanabadi, E. Sadeghi, P. Karimi, Y. Zou, E. Toyserkani, Additive manufacturing of refractory high-entropy alloys: A critical review of fundamentals and advances, Mater. Des. 260 (2025) 115261. https://doi.org/10.1016/J.MATDES.2025.115261.

[16] J. Daza, A. Wederni, R. Ullah, J. Saurina, L. Escoda, J.J. Suñol, High-Entropy Alloys Produced by Mechanical Alloying: A Review, Materials 2026, Vol. 19, Page 1300 19 (2026) 1300. https://doi.org/10.3390/MA19071300.

[17] Y. Zhang, Z. Yang, X. Mao, Z. Liu, J. Chen, Process optimization of Nb521 powder via radio-frequency plasma spherodization and its suitability for selective laser melting, Int. J. Refract. Metals Hard Mater. 138 (2026) 107742. https://doi.org/10.1016/J.IJRMHM.2026.107742.

[18] M. Kumaravel, J. Gussone, M. Sikder, A. Ardeshiri Lordejani, I.S. Berglund, J. Haubrich, S. Bagherifard, M. Guagliano, Mitigating the challenges in additive manufacturing of refractory complex concentrated alloys using Ti alloying via multi-material feedstock approach, Materials Science and Engineering: A 955 (2026) 149867. https://doi.org/10.1016/J.MSEA.2026.149867.

[19] K.S. Sista, A.P. Moon, G.R. Sinha, B.M. Pirjade, S. Dwarapudi, Spherical metal powders through RF plasma spherodization, Powder Technol. 400 (2022) 117225. https://doi.org/10.1016/J.POWTEC.2022.117225.

[20] J. Klecka, J. Cizek, J. Matejicek, F. Lukac, R. Zlatnik, T. Chraska, Tailoring the structure of RF-ICP tungsten coatings, Surf. Coat. Technol. 406 (2021) 126745. https://doi.org/10.1016/J.SURFCOAT.2020.126745.

[21] M.H. Sehhat, J. Chandler, Z. Yates, A review on ICP powder plasma spheroidization process parameters, Int. J. Refract. Metals Hard Mater. 103 (2022) 105764. https://doi.org/10.1016/J.IJRMHM.2021.105764.

[22] M. Xia, Y. Chen, K. Chen, Y. Tong, X. Liang, B. Shen, Synthesis of WTaMoNbZr refractory high-entropy alloy powder by plasma spheroidization process for additive manufacturing, J. Alloys Compd. 917 (2022) 165501. https://doi.org/10.1016/J.JALLCOM.2022.165501.

[23] X. Cui, J. Xu, J. Chen, S. Yu, J. Li, A. Akhmad, Z. Nong, Review: radio-frequency plasma spheroidization for high-sphericity powders—raw material characteristics and process parameter control, Journal of Materials Science 2026 61:35 61 (2026) 25849–25877. https://doi.org/10.1007/S10853-026-13200-1.

[24] M. Xia, Y. Chen, K. Chen, Y. Tong, X. Liang, B. Shen, Synthesis of WTaMoNbZr refractory high-entropy alloy powder by plasma spheroidization process for additive manufacturing, J. Alloys Compd. 917 (2022) 165501. https://doi.org/10.1016/j.jallcom.2022.165501.

[25] S. Gao, A. Fu, Z. Xie, T. Liao, Y. Cao, B. Liu, Preparation and Microstructure of High-Activity Spherical TaNbTiZr Refractory High-Entropy Alloy Powders, Materials 2023, Vol. 16, Page 791 16 (2023) 791. https://doi.org/10.3390/MA16020791.

[26] Y. Li, Y. Sui, Y. Feng, Y. Zhang, Y. Li, M. Song, S. Gong, Y. Xie, Powder Synthesis and Characterization of Al0.5CoCrFeNi High-Entropy Alloy for Additive Manufacturing Prepared by the Plasma Rotating Electrode Process, ACS Omega 9 (2024) 18358–18365. https://doi.org/10.1021/acsomega.4c00291.

[27] B. Storr, S.A. Catledge, High entropy alloy MoNbTaVW synthesized by metal-oxide reduction in a microwave plasma, Appl. Phys. Lett. 124 (2024) 101905. https://doi.org/10.1063/5.0192076/3270233.

[28] A. Yarlapati, Y.N. Aditya, D. Kumar, R.J. Vikram, M.K. Yadav, K.S. Reddy, K.G. Prashanth, Recent advances in additive manufacturing of refractory high entropy alloys (RHEAs): A critical review, Journal of Alloys and Metallurgical Systems 8 (2024) 100120. https://doi.org/10.1016/j.jalmes.2024.100120.

[29] Q. Shi, Y. Zhao, F. Qin, S. Peng, C. Tan, H. Xie, Y. Zhao, X. Liu, G. Zhou, Preparation of spherical WMoNbTa refractory high entropy alloy powder by radio frequency plasma and its laser powder bed fusion densification, Journal of Materials Research and Technology 38 (2025) 1539–1551. https://doi.org/10.1016/J.JMRT.2025.08.017.

[30] P. Gu, T. Qi, L. Chen, T. Ge, X. Ren, Manufacturing and analysis of VNbMoTaW refractory high-entropy alloy fabricated by selective laser melting, Int. J. Refract. Metals Hard Mater. 105 (2022) 105834. https://doi.org/10.1016/J.IJRMHM.2022.105834.

[31] P. Martin, C. Aguilar, J.M. Cabrera, A review on mechanical alloying and spark plasma sintering of refractory high-entropy alloys: Challenges, microstructures, and mechanical behavior, Journal of Materials Research and Technology 30 (2024) 1900–1928. https://doi.org/10.1016/j.jmrt.2024.03.205.

[32] M.S. Tolentino, A.G.D. Custodio, G.C. Saha, C. Aranas, Hetero-deformation-induced hardening of AlNbTaTiV refractory high-entropy alloy particles through mechanical alloying, Int. J. Refract. Metals Hard Mater. 135 (2026) 107532. https://doi.org/10.1016/J.IJRMHM.2025.107532.

[33] I. Moravcik, A. Kubicek, L. Moravcikova-Gouvea, O. Adam, V. Kana, V. Pouchly, A. Zadera, I. Dlouhy, The Origins of High-Entropy Alloy Contamination Induced by Mechanical

Alloying and Sintering, Metals 2020, Vol. 10, Page 1186 10 (2020) 1186. https://doi.org/10.3390/MET10091186.

[34] F. Lukáč, M. Vilémová, M. Klementová, P. Minárik, T. Chráska, The origin and the effect of the fcc phase in sintered HfNbTaTiZr, Mater. Lett. 286 (2021) 129224. https://doi.org/10.1016/J.MATLET.2020.129224.

[35] M. Akmal, H.K. Park, H.J. Ryu, Plasma spheroidized MoNbTaTiZr high entropy alloy showing improved plasticity, Mater. Chem. Phys. 273 (2021) 125060. https://doi.org/10.1016/J.MATCHEMPHYS.2021.125060.

[36] R. Feng, C. Tan, Z. Zhou, Y. Su, F. Gizthofer, Z. Zhou, J. Wen, S. Wen, H. Zhu, X. Liu, Fabrication of lightweight TiNbCrAlVMoZr multi-principal element alloy powder by RF plasma spheroidization, Powder Technol. 475 (2026) 122311. https://doi.org/10.1016/J.POWTEC.2026.122311.

[37] L. Xiong, A.C. Chuang, J. Thomas, T. Prost, E. White, I. Anderson, D. Singh, Defect and satellite characteristics of additive manufacturing metal powders, Advanced Powder Technology 33 (2022) 103486. https://doi.org/10.1016/j.apt.2022.103486.

[38] B. Mondal, T. Mukherjee, N.W. Finch, A. Saha, M.Z. Gao, T.A. Palmer, T. DebRoy, Vapor Pressure versus Temperature Relations of Common Elements, Materials 2023, Vol. 16, Page 50 16 (2022) 50. https://doi.org/10.3390/MA16010050.

[39] J. Brodie, J. Wang, J.P. Couzinié, M. Heczko, V. Mazánová, M.J. Mills, M. Ghazisaeidi, Stability of the B2 phase in refractory high entropy alloys containing aluminum, Acta Mater. 268 (2024) 119745. https://doi.org/10.1016/J.ACTAMAT.2024.119745.

[40] V. Soni, B. Gwalani, O.N. Senkov, B. Viswanathan, T. Alam, D.B. Miracle, R. Banerjee, Phase stability as a function of temperature in a refractory high-entropy alloy, J. Mater. Res. 33 (2018) 3235–3246. https://doi.org/10.1557/JMR.2018.223.

[41] J. Qian, C.Y. Wu, H.R. Gong, S.F. Zhou, Cohesion properties of W-ZrC interfaces from first principles calculation, J. Alloys Compd. 768 (2018) 387–391. https://doi.org/10.1016/J.JALLCOM.2018.07.238.

[42] R. Hu, S. Jin, G. Sha, Application of atom probe tomography in understanding high entropy alloys: 3D local chemical compositions in atomic scale analysis, Prog. Mater. Sci. 123 (2022) 100854. https://doi.org/10.1016/J.PMATSCI.2021.100854.

[43] M. Vaidya, A. Karati, A. Marshal, K.G. Pradeep, B.S. Murty, Phase evolution and stability of nanocrystalline CoCrFeNi and CoCrFeMnNi high entropy alloys, J. Alloys Compd. 770 (2019) 1004–1015. https://doi.org/10.1016/J.JALLCOM.2018.08.200.

[44] M. Vaidya, A. Karati, K. Guruvidyathri, M. Nagini, K.G. Pradeep, B.S. Murty, Suppression of σ-phase in nanocrystalline CoCrFeMnNiV high entropy alloy by unsolicited contamination during mechanical alloying and spark plasma sintering, Mater. Chem. Phys. 255 (2020) 123558. https://doi.org/10.1016/J.MATCHEMPHYS.2020.123558.

[45] R. John, A. Karati, M.M. Garlapati, M. Vaidya, R. Bhattacharya, D. Fabijanic, B.S. Murty, Influence of mechanically activated annealing on phase evolution in Al0.3CoCrFeNi high-entropy alloy, Journal of Materials Science 2019 54:23 54 (2019) 14588–14598. https://doi.org/10.1007/S10853-019-03917-7.

[46] J. Umeda, H. Ishizaka, S. Li, A. Alhazaa, K. Kondoh, Comparison study on mechanical properties of powder metallurgy titanium materials with nitrogen solutes and TiN dispersoids, J. Alloys Compd. 846 (2020) 156455. https://doi.org/10.1016/J.JALLCOM.2020.156455.

[47] N.Y. Yurchenko, N.D. Stepanov, S. V. Zherebtsov, M.A. Tikhonovsky, G.A. Salishchev, Structure and mechanical properties of B2 ordered refractory AlNbTiVZrx (x = 0–1.5) high-entropy alloys, Materials Science and Engineering: A 704 (2017) 82–90. https://doi.org/10.1016/J.MSEA.2017.08.019.

[48] J. Kozlík, F. Lukáč, M. Casas-Luna, J. Veselý, E. Jača, K. Ficková, S. Šašek, K. Bartha, A. Strnad, T. Chráska, J. Stráský, Phase equilibria of the Al–Ti–Nb–Zr–Ta system, J. Mater. Sci. (2026). https://doi.org/10.1007/s10853-026-13083-2.

[49] T.E. Whitfield, E.J. Pickering, L.R. Owen, O.N. Senkov, D.B. Miracle, H.J. Stone, N.G. Jones, An assessment of the thermal stability of refractory high entropy superalloys, J. Alloys Compd. 857 (2021) 157583. https://doi.org/10.1016/J.JALLCOM.2020.157583.